%% file: main.tex
\documentclass[conference]{IEEEtran}
\IEEEoverridecommandlockouts

\newif\ifEditMode
\EditModetrue

\usepackage{algorithm2e}
\RestyleAlgo{ruled}

\usepackage{cite}
\usepackage{amsmath,amssymb,amsfonts}
\usepackage{graphicx}
\usepackage{textcomp}
\usepackage{xcolor}
\usepackage{aliases}
\usepackage{preamble}
\usepackage{float}

\usepackage{amsthm}

\def\BibTeX{{\rm B\kern-.05em{\sc i\kern-.025em b}\kern-.08em
    T\kern-.1667em\lower.7ex\hbox{E}\kern-.125emX}}
\begin{document}

\title{On Exercising Governance Power in Decentralized Autonomous Organizations}

\author{\IEEEauthorblockN{Vabuk Pahari}
\IEEEauthorblockA{
\textit{Max Planck Institute for Software Systems}\\
 Germany \\
vpahari@mpi-sws.org}
\and
\IEEEauthorblockN{Balakrishnan Chandrasekaran}
\IEEEauthorblockA{
\textit{Vrije Universiteit Amsterdam}\\
Amsterdam, Netherlands \\
b.chandrasekaran@vu.nl}
\and
\IEEEauthorblockN{Johnnatan Messias}
\IEEEauthorblockA{
\textit{Max Planck Institute for Software Systems}\\
 Germany \\
johnme@mpi-sws.org}
\and
\IEEEauthorblockN{Krishna P. Gummadi}
\IEEEauthorblockA{
\textit{Max Planck Institute for Software Systems}\\
 Germany \\
gummadi@mpi-sws.org}
\and
\IEEEauthorblockN{Abhisek Dash}
\IEEEauthorblockA{
\textit{Max Planck Institute for Software Systems}\\
 Germany \\
adash@mpi-sws.org}
}

\maketitle

\input{00-abstract}
\input{01-introduction}

\input{table-daos-removed}
\input{02-dataset-methodology}

\input{04-how-is-power-exercised}

\input{05-real-world-attacks}

\input{06-related-work}

\input{07-conclusion}

\bibliographystyle{IEEEtran}
\bibliography{references}

\appendix

\input{appendix}

\end{document}

%% file: 00-abstract.tex
\begin{abstract}

A decentralized autonomous organization (DAO) is a governance entity that allows its stakeholders to manage blockchain-based protocols through smart contracts.
The DAO explicitly specifies how stakeholders make and enforce decisions concerning a protocol's operation in a smart contract, aptly referred to as its \emph{governance contract}.
The design of this governance contract, therefore, has far-reaching implications for the security (trust) and privacy (transparency) of the smart contracts managed by the DAO and its stakeholders.
In this work, we (i) explicate the trust and transparency trade-offs of the design choices in implementing a DAO and (ii) highlight how poor choices introduce critical vulnerabilities, using real-world examples as case studies.
To this end, we analyze $48$ public, actively used Ethereum-based DAOs that control a vast capital.
We classify the design choices into a handful of key dimensions that succinctly capture how a DAO's stakeholders initiate a protocol change, vote on it, and, based on the voting outcome, execute that change.
Our analyses crucially uncover a new class of attacks, which we call \emph{governance attacks}, that directly exploit the fundamental design of a DAO's governance mechanisms, even if we assume bug-free implementations.

\end{abstract}


%% file: 01-introduction.tex
\section{Introduction}\label{s:introduction}

\Ac{defi} represents a major shift away from traditional financial systems by eliminating intermediaries in key sectors such as banking, lending, and insurance~\cite{adams2021uniswap, Daian@S&P20, Qin@FC21, Perez@FC21}.
\ac{defi} protocols instead rely on open-source smart contracts deployed on public, permission-less blockchains, which enable transparent, algorithmic interactions among global participants.
They emerged around 2017 and have grown rapidly since then, securing more than 150 billion USD in \ac{tvl}~\cite{defillama} today.
The sustainability of this \ac{defi} model depends critically on how the underlying smart contracts are governed (i.e., managed and updated).

Today, the governance of \ac{defi} smart contracts is facilitated by \Acp{dao}.
\Acp{dao} offer well-specified mechanisms---encoded in smart (“governance”) contracts---for their stakeholders to propose protocol changes, vote on such proposals, and, upon succeeding, enact the proposed changes.
Stakeholders' voting rights and powers are typically determined based on the volume of governance tokens that they hold or have “invested” in the \ac{dao}.
Since the governance tokens are \stress{publicly} traded---anyone can buy these tokens, anonymously, and acquire governance power---the manner in which governance power is exercised by a \ac{dao}'s stakeholders has far-reaching trust and transparency concerns for the \ac{dao} and its stakeholders.
A \ac{dao}'s governance contract, hence, unsurprisingly, adopts various safeguards to protect itself from attackers.

A \ac{dao}'s safeguards range from “gatekeeping” stakeholders', by controlling how and when they can exercise their governance power, to introducing “guard rails” for the decentralized decision making process, by empowering (centralized) authorities to control (e.g., veto) the voting outcomes, if required.
Although these safeguards are crucial for making the \acp{dao} resilient to attacks, they violate the \acp{dao}' fundamental design tenet---decentralized governance.
In this work, we investigate such \ac{dao} safeguards to expose the design choices and trade-offs involved in their implementations.

We analyzed the governance contracts underlying \surveysz{} \acp{dao} with the largest treasury sizes~\cite{deepdao}, reviewed their software implementation, and read their developer documentations and specifications.
Though virtually all of governance contracts differ from one another in some respects, we identified a set of key dimensions along which their designs differ.
These design choices reveal the complex trade-offs that designers of governance contract need to balance between realizing the decentralized self-governance ideals of \ac{dao} (in theory) and improving trust and transparency in governance (in practice).
We present a two-fold analyses of these design choices and trade-offs.
First, we analyze the formal governance mechanisms, specifically the process by which a referendum is initiated, voted on, certified, and executed in a \ac{dao}.
Then, we discuss real-world governance attacks against \acp{dao} and highlight how they succeeded (or failed) due to the choices made by the \ac{dao} designers.
We summarize our key contributions as follows.

\point{}
We review the (software) implementations of the smart contracts underlying
\surveysz{} \acp{dao} and peruse through developer documentations to identify the key
admission controls and guard rails used in governance protocols (\autoref{tab:gov-proto-survey-full}).

\point{}
We highlight the design choices in implementing these mechanisms throughout the
entire life cycle of a referendum, and we analyze their implications for the
trust and transparency of the \acp{dao}.
We find, for instance, that while the vetoer mechanism enhances a \ac{dao}'s security and delegation improves its efficiency, they both cripple the promise of decentralized governance.

\point{}
We discover a new class of attacks on \acp{dao} and call them \stress{governance attacks}.
We show that the vulnerabilities exploited in recent attacks persist among many of the \acp{dao} we investigated because of fundamental (mechanism) design issues.
Barring changing the design of their safeguards, these implementations can be “patched” to render these \acp{dao} resilient to governance attacks.

\point{}
We will release the data we gathered and our code for analyzing the data
as open-source artifacts upon publication.

%% file: table-daos-removed.tex
\begin{table*}[btp]
\tabcap{%
Our investigation of \surveysz{} DAOs and their governance structures.
“ID” refers to the assigned id for the DAO, which should be used by the reader when identifying DAOs in the paper.
`\textbf{Call for Referendum (Call)}' is one of `Voter Initiated (\callVI)', `Authority Initiated (\callAI)', or Anybody (\callAny).
\textbf{Voting platform (`VP')} is either on-chain (\onchain) or off-chain (\offchain).
`\textbf{Voting Format (VF)}' can be `Periodic Voting (\fmtPerd{})', `Continuous Polling (\fmtCont{})',
`\textbf{Voting Type (VT)}' can be `Proactive (P)', `Reactive (R)', 
`\textbf{Voting Aggregation (VA)}' can be `Vote Differential (V)', `Majority (M)', `Threshold (T)', or `Plurality (P)'.
`\textbf{Certification (Cert.)}' is one of \{`Centralized (\centralized)', `Optimistic Certification (\optimistic)', Anybody (\callAny)\}.
`\textbf{Execution (Exec.)} is one of \{`Centralized (\centralized)', `Anybody (\callAny)'\}.
`\textbf{Veto (Veto)}' is one of \{`Vote Initiator (I)', `Veto Authority (A)', `Nobody (–)' \}.}\label{tab:gov-proto-survey-full}%
\centering
\resizebox{0.8\textwidth}{!}{%
\begin{tabular}{@{}rrcccccccc@{}}
\toprule
\thead{ID} & \thead{Protocol} & \thead{Call} & \thead{VP} & \thead{VF} & \thead{VT} & \thead{VA} & \thead{Cert.} & \thead{Exec.} & \thead{Veto} \\
\midrule

1 & \textit{Uniswap}~\cite{Governance@Uniswap} & \callVI{} & \onchain & \fmtPerd{} & P & M & \callAny & \callAny & I \\

2 & \textit{ENS}~\cite{ens@developerdocs} & \callVI{} & \onchain & \fmtPerd{} & P & M & \callAny & \callAny & A \\

3 & \textit{Maker}~\cite{Governance@MakerDAO} & \callAny{} & \onchain & \fmtCont{} & P & P & \callAny & \callAny & – \\

4 & \textit{Lido (Easy Track)}~\cite{lido@easytrack} & \callAI{} & \onchain & \fmtPerd{} & R & M & \callAny & \callAny & I, A \\

4 & \textit{Lido Governance}~\cite{lido@developerdocs} & \callVI{} & \onchain & \fmtPerd{} & P & M & \callAny & \callAny & – \\

5 & \textit{Frax Finance Omega}~\cite{fraxfinance@developerdocs} & \callAI{} & \onchain & \fmtPerd{} & R & M & \callAny & \callAny & A \\

5 & \textit{Frax Finance Alpha}~\cite{fraxfinance@developerdocs} & \callVI{} & \onchain & \fmtPerd{} & R & M & \callAny & \callAny & – \\

6 & \textit{AAVE}~\cite{Governance@AAVE} & \callVI{} & \onchain & \fmtPerd{} & P & V & \callAny & \callAny & A \\

7 & \textit{Compound}~\cite{Governance@Compound} & \callAI{}\,\callVI{} & \onchain & \fmtPerd{} & P & M & \callAny & \callAny & I, A \\

8 & \textit{Radicle}~\cite{radicle@developerdocs} & \callVI{} & \onchain & \fmtPerd{} & P & M & \callAny & \callAny & – \\

9 & \textit{0x Protocol}~\cite{0x@developerdocs} & \callVI{} & \onchain & \fmtPerd{} & P & M & \callAny & \callAny & – \\

10 & \textit{Gitcoin}~\cite{Gitcoin} & \callVI{} & \onchain & \fmtPerd{} & P & M & \callAny & \callAny & – \\

11 & \textit{Silo Finance}~\cite{silofinance@developerdocs} & \callVI{} & \onchain & \fmtPerd{} & P & M & \callAny & \callAny & – \\

12 & \textit{Lyra}~\cite{lyra@developerdocs} & \callVI{} & \onchain & \fmtPerd{} & P & M & \callAny & \callAny & A \\

13 & \textit{API3}~\cite{api3@developerdocs} & \callVI{} & \onchain & \fmtPerd{} & P & M & \callAny & \callAny & – \\

14 & \textit{Ampleforth}~\cite{ampleforth@developerdocs} & \callVI{} & \onchain & \fmtPerd{} & P & M & \callAny & \callAny & I \\

15 & \textit{Instadapp}~\cite{instadapp@developerdocs} & \callVI{} & \onchain & \fmtPerd{} & P & M & \callAny & \callAny & I, A \\

16 & \textit{Rari}~\cite{rari@developerdocs} & \callVI{} & \onchain & \fmtPerd{} & P & M & \callAny & \callAny & I, A \\

17 & \textit{NounsDAO}~\cite{nounsdao@developerdocs} & \callVI{} & \onchain & \fmtPerd{} & P & M & \callAny & \callAny & I, A \\

18 & \textit{Curve}~\cite{curve@developerdocs} & \callVI{} & \onchain & \fmtPerd{} & P & T & \callAny & \callAny & – \\

19 & \textit{Origin}~\cite{origin@developerdocs} & \callVI{} & \onchain & \fmtPerd{} & P & M & \callAny & \callAny & I \\

20 & \textit{Hop DAO}~\cite{hopdao@developerdocs} & \callVI{} & \onchain & \fmtPerd{} & P & M & \callAny & \callAny & – \\

21 & \textit{Cryptex}~\cite{cryptex@developerdocs} & \callVI{} & \onchain & \fmtPerd{} & P & M & \callAny & \callAny & – \\

22 & \textit{Angle Protocol}~\cite{angle@developerdocs} & \callVI{} & \onchain & \fmtPerd{} & P & M & \callAny & \callAny & I, A \\

23 & \textit{DxDao}~\cite{dxdao@developerdocs} & \callAny{} & \onchain & \fmtPerd{} & P & T & \callAny & \callAny & – \\

24 & \textit{Nexus Mutual}~\cite{nexus@developerdocs} & \callAI{} & \onchain & \fmtPerd{} & P & M & \callAny\,\centralized & \callAny & A \\

25 & \textit{Goldfinch}~\cite{goldfinch@developerdocs} & \callAI{}\,\callVI{} & \offchain & \fmtPerd{} & P & M & \centralized & \centralized & I, A \\

26 & \textit{ParagonsDAO}~\cite{paragonsdao@developerdocs} & \callAI{} & \offchain & \fmtPerd{} & P & T & \centralized & \centralized & I, A \\

27 & \textit{Illuvium}~\cite{illuvium@developerdocs} & \callAI{} & \offchain & \fmtPerd{} & P & T & \centralized & \centralized & I, A \\

28 & \textit{SuperRare}~\cite{superrare@developerdocs} & \callAI{} & \offchain & \fmtPerd{} & P & M & \centralized\,\optimistic & \centralized & I, A \\

29 & \textit{Mantle}~\cite{mantle@developerdocs} & \callAI{} & \offchain & \fmtPerd{} & P & M & \centralized & \centralized & I, A \\

30 & \textit{Res. Hub Fdn.}~\cite{researchhub@developerdocs} & \callAI{}\,\callVI{} & \offchain & \fmtPerd{} & P & M & \centralized & \centralized & I, A \\

31 & \textit{Stargate Finance}~\cite{stargate@developerdocs} & \callAI{}\,\callVI{} & \offchain & \fmtPerd{} & P & M & \centralized & \centralized & I, A \\

32 & \textit{Uma}~\cite{uma} & \callAI{} & \offchain & \fmtPerd{} & P & M & \centralized & \centralized & I, A \\

33 & \textit{Cowswap}~\cite{cowprotocol@developerdocs} & \callAI{}\,\callVI{} & \offchain & \fmtPerd{} & P & M & \centralized\,\optimistic & \centralized & I, A \\

34 & \textit{Sturdy Finance}~\cite{sturdy@developerdocs} & \callAI{}\,\callVI{} & \offchain & \fmtPerd{} & P & M & \centralized & \centralized & I, A \\

35 & \textit{Euler}~\cite{euler@developerdocs} & \callAI{}\,\callVI{} & \offchain & \fmtPerd{} & P & M & \centralized & \centralized & I, A \\

36 & \textit{SAFE}~\cite{safe@developerdocs} & \callAI{}\,\callVI{} & \offchain & \fmtPerd{} & P & M & \centralized & \centralized & I, A \\

37 & \textit{Tokenlon}~\cite{tokenlon@developerdocs} & \callAI{} & \offchain & \fmtPerd{} & P & M & \centralized & \centralized & I, A \\

38 & \textit{Botto}~\cite{botto@developerdocs} & \callAI{} & \offchain & \fmtPerd{} & P & M & \centralized & \centralized & I, A \\

39 & \textit{Balancer}~\cite{balancer@developerdocs} & \callAI{}\,\callVI{} & \offchain & \fmtPerd{} & P & M & \centralized & \centralized & I, A \\

40 & \textit{Sushiswap}~\cite{sushiswap@developerdocs} & \callAI{} & \offchain & \fmtPerd{} & P & M & \centralized & \centralized & I, A \\

41 & \textit{Gearbox}~\cite{gearbox@developerdocs} & \callAI{}\,\callVI{} & \offchain & \fmtPerd{} & P & M & \centralized & \centralized & I, A \\

42 & \textit{Paraswap}~\cite{paraswap@developerdocs} & \callAI{}\,\callVI{} & \offchain & \fmtPerd{} & P & M & \centralized & \centralized & I, A \\

43 & \textit{Alchemix}~\cite{alchemix@developerdocs} & \callAI{} & \offchain & \fmtPerd{} & P & M & \centralized & \centralized & I, A \\

44 & \textit{1Inch}~\cite{1inch@developerdocs} & \callAI{}\,\callVI{} & \offchain & \fmtPerd{} & P & M & \centralized & \centralized & I, A \\

45 & \textit{Shutter DAO 0x36}~\cite{shutter@developerdocs} & \callAI{}\,\callVI{} & \offchain & \fmtPerd{} & P & M & \centralized & \centralized & I, A \\

46 & \textit{Yearn Finance}~\cite{yearn@developerdocs} & \callAI{}\,\callVI{} & \offchain & \fmtPerd{} & P & M & \centralized & \centralized & I, A \\

47 & \textit{Shapeshift}~\cite{shapeshit@developerdocs} & \callAI{}\,\callVI{} & \offchain & \fmtPerd{} & P & M & \centralized & \centralized & I, A \\

48 & \textit{Decentraland}~\cite{decentraland@developerdocs} & \callAI{} & \offchain & \fmtPerd{} & P & M & \centralized & \centralized & I, A \\

\bottomrule
\end{tabular}%
}

\end{table*} 

%% file: 02-dataset-methodology.tex
\begin{figure*}[tb]
  \centering
  \includegraphics[width=0.75\textwidth]{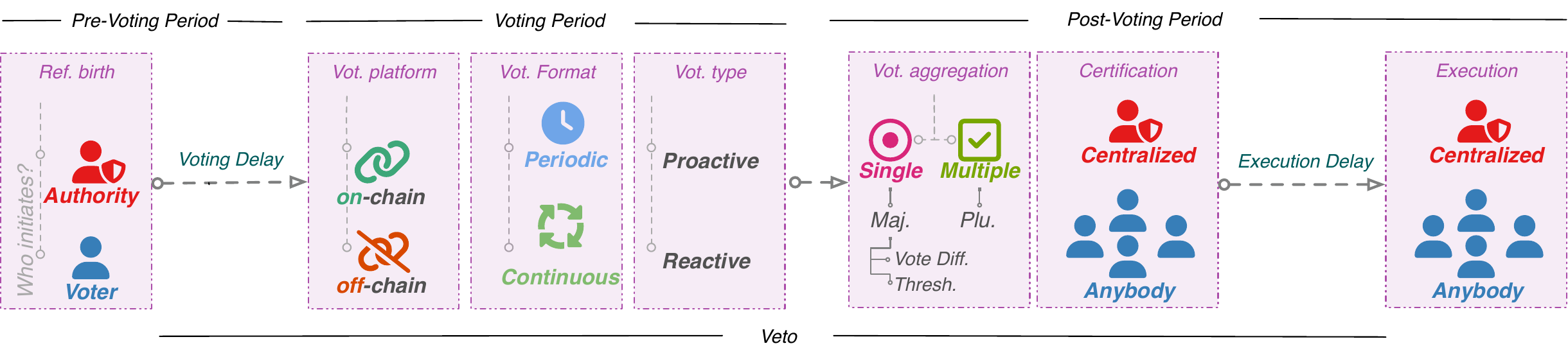}
  \figcap{Lifecycle of a referendum involving its birth, being put to vote, ratification of the voting, and, if approved, its execution.}
  \label{fig:voting-parameter}
\end{figure*}

\section{Methodology}\label{s:background}

We selected $48$ \acp{dao} for investigation as follows.
First, we focused on \acp{dao} whose governance contracts are deployed on Ethereum, the blockchain with the highest on-chain \ac{tvl}~\cite{defillama}.
Second, we selected \acp{dao} with the largest treasury sizes, as reported by DeepDAO in May 2024~\cite{deepdao}.
Many of these \acp{dao} correspond to leading \ac{defi} projects by \ac{tvl}, indicating substantial user trust and influence within the \ac{defi} ecosystem. 
Third, we excluded \stress{private} \acp{dao} (e.g., the Graph Protocol\cite{graph@developerdocs}), where decision-making authority is limited only to a small subset of stakeholders, a model that is fundamentally at odds with decentralization.
Our study, hence, consists of \stress{public} \acp{dao}, where anybody can participate in governance either by buying governance tokens in the market or doing some work for the \ac{dao}.
Finally, we excluded \acp{dao} that had not conducted a governance vote in the six months preceding May 2024, thereby removing inactive projects with potentially outdated documentation or discontinued operations.

For each of the \surveysz{} \acp{dao} selected, we parsed the official documentations to fetch contract addresses and their off-chain voting platforms.
Then, we gathered all their governance contracts and governance-token contracts using Etherscan.
\acp{dao} publish the relevant contract code on Etherscan, and Etherscan validates the correctness of the code by compiling and checking that its bytecode matches the bytecode of that deployed on Ethereum.
We also ran an Ethereum \stress{archive} node using the Erigon Client to gather data from these contracts.
Lastly, we collected off-chain voting data of \acp{dao} using the official Snapshot API~\cite{Snapshot,Snapshot-api}.

%% file: 04-how-is-power-exercised.tex
\section{Exercising Governance Power}\label{s:voting-parameters}

The fundamental objective of a DAO is to facilitate its stakeholders to enact changes to it and manage its finances in a decentralized manner.
These tasks are accomplished through (change) proposals or \newterm{referendums}, which then solicit the votes (or approvals) of stakeholders, and the voting outcomes shape the protocol's evolution.
In~\autoref{fig:voting-parameter}, we enumerate the lifecycle of a referendum by dividing it into three broad temporal segments: (a)~\stress{pre-voting period}, (b)~\stress{voting period}, and (c)~\stress{post-voting period}.
In this section, we elaborate on these temporal segments and discuss the implications of the design choices associated with each segment.

\subsection{Pre-Voting Period}

This period starts from before a referendum is initiated and ends when the voting begins.

\subsubsection{The birth of a referendum}\label{s:proposer}
The right to initiate a referendum is typically bestowed on only a subset of the stakeholders of a DAO, e.g., founders, investors, or users with substantial stake in the protocol.
Constraining this right effectively protects a DAO from being spammed by random stakeholders with arbitrary proposals.
The stakeholder who initiates a referendum is referred to as the \newterm{proposer}, and their type can be used to classify referenda.

\paraib{Anybody}
In this approach, 
\stress{anyone}  can initiate a referendum, regardless of their governance power in the DAO.
Only Maker and Dxdao (\#3 and \#23~in~\autoref{tab:gov-proto-survey-full}).

\paraib{Voter-initiated}\label{s:proposal-threshold}
DAOs using this approach only allow voters (or stakeholders) with some minimum voting power, called \newterm{proposal threshold}, to initiate a referendum.
A voter-initiated referendum is akin to a ballot initiative in California, US~\cite{californiaballotinitiative};
to be deemed eligible for voting, the initiative must garner a threshold number of signatures from the state's residents.
The proposer must have a minimum threshold of voting power.
Most protocols also stipulate that the proposer must maintain this threshold voting power until the referendum is enacted, failing which the referendum can be immediately canceled.
In our study 19 DAOs use \stress{only} voter-initiated referenda.

\paraib{Authority-initiated}\label{s:proposer-role}
Here, only certain \stress{trusted} members (identified through their wallet addresses) are allowed to be proposers.
This approach is analogous to how only the monarch has the legal authority to call a general election in the UK~\cite{generalelection}.
In our study, 13 DAOs use \stress{only} authority-initiated referenda, and all DAOs with off-chain voting allow for authority-initiated referenda.

\begin{mdframed}[style=Takeaways]
\takeaways{Trust and Transparency trade-offs}
If anyone can initiate a referendum, the DAO can be spammed by random proposals.
People that do not have any stake in the DAO can initiate malicious proposals, thereby undermining the security of the DAO.
Dxdao (\#23), hence, uses \stress{non-transferable} governance tokens, and it also has a decentralized voter distribution---the largest voter in Dxdao has about 5\% of the total voting power~\cite{dxdaotokendistribtuion}.
Maker (\#3), in contrast, uses a centralized process to address the security concerns: Only the `core' team of Maker are involved in the process, and Maker's governance portal only displays proposals that are put forward by these governance \newterm{facilitators}~\cite{govfacilitator@makerdao}.

Proposal thresholds in voter-initiated referenda serve as an effective ``guard rail'' to limit, but not eliminate spam proposals~\cite{feichtinger@sokattacksondao}.
Mandating that a proposer maintain the minimum voting power until the proposal is executed, however, reduces nefarious referenda, since the proposer themselves must face the outcomes of the referendum.

Authority-initiated referendas only allow a select group of stakeholders to initiate a referendum.
Even if other users with substantial economic stakes or the majority of stakeholders support a specific change, they cannot initiate a referendum unless they belong to this select group.
As a consequence, the decision-making process becomes centralized to this select group; a voter-initiated referenda, in contrast, fosters a diverse pool of proposers.

\end{mdframed}

\subsubsection{Voting Delay}~\label{s:voting-delay}
DAOs often impose a voting delay between initiating a proposal and starting a vote to allow time for community support or opposition, especially if no prior discussions occur.
A short delay might give malicious actors a chance to buy tokens and propose a referendum within the same block, leaving little time for others to react.
11 DAOs in our study have voting delays of 1 to 7 days (for 0x (\#9)), while 8 have no delay. For example, Lido and Maker allow immediate voting, and ENS, Radicle, and Cryptex start voting in the next block. Most DAOs with off-chain voting (except 4) don’t specify a delay.

\subsubsection{Pre-proposal Opinion Polling}

Some protocols, like Uniswap, AAVE, and Gitcoin (\#1, \#6 and 10, resp.), use polling as a signaling tool before the official vote. 
Polling gauges community support for a proposal at no cost, and if approved, the proposal moves to a binding vote. 
However, this step is a procedural norm, and a user can bypass it to directly initiate an official referendum, which can pass and be executed regardless of polling results.

\subsection{Voting Period}\label{s:voting-period}

The voting period is the time when votes can be cast; outside this period, votes are invalid. 
Some DAOs have a rigid voting period, which cannot be changed, making them susceptible to \stress{vote sniping} \cite{feichtinger@sokattacksondao}, where a voter can delay their votes until the end of the voting period, thereby deciding whether a referendum is passed in the final second.
To circumvent such attacks Angle, Origin and Frax extends the voting period by 2 days, if the quorum, i.e. minimum number of votes required for a referendum to
pass, was reached late. Lido only allows users to vote against a proposal on the last day of voting.
This disincentivizes users from potentially waiting until the last minute to vote and influence the outcome.

In all DAOs, except Illuvium, every user's vote is seen soon as it is cast, and there is a running tally of the vote count.
If a voter sees that their desired choice is winning, then they might not feel obliged to cast their vote, especially because voting incurs a cost \cite{austgen2023dao}.
Illuvium, uses \stress{Shielded Voting}\cite{shielded-voting}, where votes are encrypted and results remains private until the end of the voting period. The voting here is, however, done off-chain, and such a scheme can be expensive to implement on-chain.
There is, however, no voter privacy after the election period, as every user's vote is revealed after the election.
Austgen \ea show that private voting reduces the centralizing effect of \stress{herding}~\cite{austgen2023dao}, where users vote similarly to some influential members of the DAO \cite{sharma2023unpacking}.

\subsubsection{Voting Format} In our study, only Maker (\# 3), which uses Continuous (approval) voting, does not have an explicit voting period.
Continuous voting~\cite{Governance@MakerDAO}, requires a referendum to maintain more votes than any other (alternative) referendum to be deemed accepted.
A referendum can be accepted at any time in the future, if it receives more votes than the current leading referendum. The same referendum can even be accepted twice. For example, if enough voters move their votes from an old referendum, to a new referendum and back to the old, then the same referendum is accepted twice; but some parameters in a referendum can only be executed once.

\subsubsection{Voting Type} 
In all DAOs except for Lido and Frax, a referendum is ``proactive'', meaning that a \stress{quorum} of votes must be in favor for the proposal to pass. 
A proactive referendum calls for users to record their explicit consent to change the protocol.
Since voting power can be bought, quorums serve as a safeguard by reducing the likelihood of such power transfers from influencing the outcome of a referendum.
Quorums also prevent proposals with low engagement from passing, ensuring only changes approved by a significant number of users.

In contrast, ``reactive'' referendums (or \stress{optimistic} voting) are considered to have been passed, unless enough votes are cast against them.
These also require a quorum; however, the quorum is voting on preventing, rather than promoting, a change from happening.
Since a proposal by default effects a change, reactive referenda can only be initiated by some authority (recall \S\ref{s:proposer}).
Otherwise, malicious actors can enervate the community by continually requiring them to vet proposals that harm or destabilize the protocol.

\subsubsection{Voting Platform}\label{s:voting-platform}

Once a referendum is created, stakeholders vote on the referendum to either pass or reject it.
A voting platform offers an avenue for the stakeholders to cast their votes, and governance protocols have two platforms, \newterm{on-chain} and \newterm{off-chain} voting, at their disposal.
Even in the real world, various voting platforms exist (e.g., in-person, electronic, and mail-in voting), each with its own accuracy, integrity, and accessibility implications.
The electronic voting platform, for instance, can foster greater voter turnout in elections compared to in-person voting, but it introduces other challenges such as voting system security and voter's susceptibility to coercion \cite{rosario@votercoercion, merino2024evote}.
Below, we compare and contrast the properties of the two voting platforms employed by DAOs.

\paraib{On-chain voting}\label{s:on-chain-voting}
In this platform, stakeholders cast their votes directly on a blockchain, ensuring transparency and immutability.
Smart contracts can automatically execute protocol changes, depending on the voting results, adhering to governance rules encoded within the contracts.
In an on-chain voting platform, all the rules for a referendum, from proposing to execution, is encoded in the smart contract.
In our study, 24 DAOs use on-chain voting.

\paraib{Off-chain Voting.}\label{s:off-chain-voting}
Votes in this platform are recorded outside the blockchain, often using a third-party platform, e.g., Snapshot~\cite{Snapshot}.
(All DAOs in our study with off-chain voting use Snapshot.)
Snapshot allows a protocol to create a public ``space'' for discussing its governance, establish voting rules, and determine voting rights based on the on-chain state.
Although the voting happens off-chain, Snapshot still uses a user's (on-chain) wallet address(es) for authenticating their vote(s).
Votes are stored on a decentralized storage system~\cite{IPFS}.
In our study, 24 DAOs use off-chain voting.

\begin{figure}[tb]
    \centering
    \includegraphics[width=3.4in]{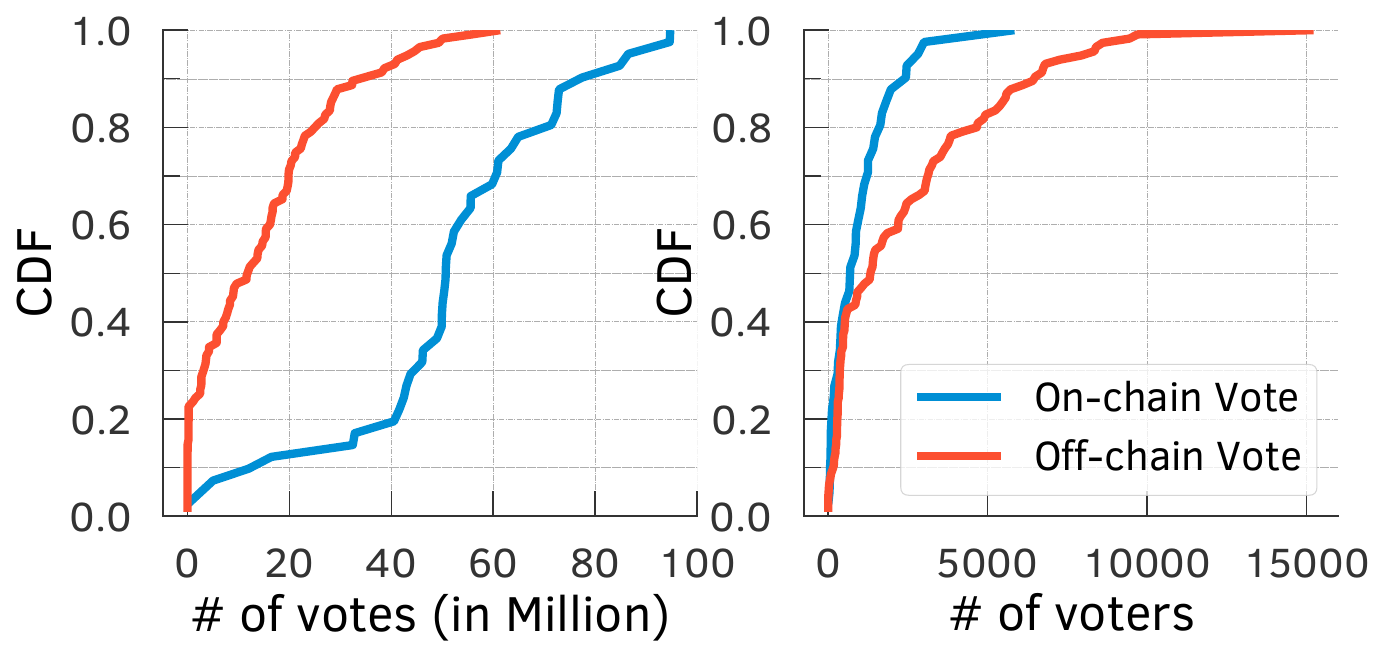}
    \figcap{Voting participation and number of votes cast in Uniswap for on-chain and off-chain voting. Procedurally, Uniswap uses an off-chain signaling vote before an official on-chain vote. We see there is a larger voter turnout in off-chain votes, possibly because it is free, allowing small users to cast their votes. But, since this is only a signaling mechanism and not official, voters with large votes might not necessarily participate. Hence, there is a much larger number of votes cast in on-chain voting, where users with large voting power are more likely to vote.}
    \label{fig:uniswap-on-chain-off-chain}
    \end{figure}

\begin{mdframed}[style=Takeaways]
\takeaways{Trust and Transparency trade-offs}
On-chain voting platforms have no additional dependencies on a third party for executing passed proposals.
The well-specified rules and automatic proposal executions of the platform can, nevertheless, allow a user to engineer malicious proposals to pass.
The Synthetify DAO was attacked, for instance, by a user who acquired enough voting power and passed a malicious proposal that transferred the protocol's treasury to themselves~\cite{feichtinger@sokattacksondao}.
The voting costs incurred by stakeholders in an on-chain voting platform may deter participation, especially among users with few voting rights (i.e., small economic stake)~\cite{galaxy@govdefi}.
Prior work has shown that the cost per vote for `small' users is significantly higher than that for larger voters, precipitating in low voter turnout and concentration of voting power among few large stakeholders~\cite{messias2023gov}.

The off-chain voting platform, in contrast, eliminates the cost barrier, enabling everyone to vote on a proposal.
The platform, unsurprisingly, leads to a larger voting participation compared to the on-chain platform~\cite{galaxy@govdefi}.
In \autoref{fig:uniswap-on-chain-off-chain}, we observe that in Uniswap, which employs on-chain voting with pre-proposal opinion polling, a larger number of voters participate in off-chain proposals, showing that voters do in fact participate more when there is no cost to voting.
Although users of this platform do not pay any poll tax to cast their vote(s), the platform requires a `trusted' third party to execute any passed proposals~\cite{zamyatin@fc}.
Unlike the on-chain voting platform, stakeholders' preferences or votes on a proposal are, hence, non-binding.
However, the platform's dependency on a third party for execution introduces a huge risk: They can choose \stress{not} to execute the proposal, particularly if they deem it malicious or harmful.

\end{mdframed}

\subsection{Post-Voting Period}\label{s:post-voting-period}

\subsubsection{Vote Aggregation}\label{s:voting-aggregation}

Vote aggregation refer to rules for how votes are tallied and the rules for determining the outcome of a proposal.
Aggregation rules can significantly impact the outcome of a proposal.
A real-world analogy is how measures are ratified in the U.S. Senate:
While most measures pass with a simple majority, a two-thirds majority is needed to ratify constitutional amendments~\cite{senatevoting}.
Common aggregation rules include majority, ranked choice, and plurality voting, each with trade-offs om fairness, efficiency, stability, etc \cite{vote-counting-schemes}.
We categorize these rules into
\stress{single} and \stress{multiple} choice.

\paraib{Single choice}
In a single-choice referendum, voters either accept or reject a proposal.
The proposal must reach a quorum, and majority of voters must vote in favor for the proposal to pass.
All DAOs with on-chain voting except for Maker opt for single-choice referenda.
Although the majority determines the outcome in single-choice referenda, there are a few key distinctions in how they are determined in practice.

\textit{(i) Vote Differential} requires a minimum difference between votes in favor and those against for a referendum to pass.
Majority voting is a simple form of vote differential where this difference is one.
Most DAOs use majority voting with only AAVE (\#6), requiring a differential of $\num{80000}$ votes.

\textit{(ii) Threshold voting} requires a threshold number of votes in favor for a referendum to pass.
This method is used by councils, where minimum number of members need to vote in favor for a proposal to pass.
Furthermore, Curve (\#18) requires a specific threshold of $51\%$ of votes for a proposal to pass instead of a simple majority.

\paraib{Multiple Choice}
Instead of voting for or against a specific change to the protocol, here, voters declare their preferences towards two or more choices. 
Off-chain voting protocols allow for multiple choice, albeit most votes are single choice \cite{empiricalsnapshot}.

\textit{Plurality voting}:
Maker is the only DAO with on-chain voting that uses plurality voting.
The DAO accepts a new proposal if it receives more votes than the current proposal (refer \S\ref{s:voting-period}). 
The winning proposal does not need to receive a majority of votes, but only more than any of the other alternatives, although in practice they do so~\cite{makerdaopeculiarity}.

\subsubsection{Certification and Execution}\label{sec:decentralized-cerifier-executor}
After a proposal passes, it is certified and executed. 
The former entails checking the criteria for passing a proposal, while the latter constitutes the actual execution of the proposal.
These privileges can be are assigned to a specific authority, with exclusive rights, similar to how the U.S. Congress certifies the results of the presidential election by officially counting the electoral votes~\cite{ussenatecertification}.
In on-chain protocols, certification and execution are typically written into the smart contract(s) underlying the DAO.

\paraib{Anybody}
In DAOs with on-chain voting, \stress{anybody} can invoke the relevant smart contract function to certify a referendum.
The function, upon invocation, checks if the referendum meets the (pre-defined) vote aggregation rules (\S\ref{s:voting-aggregation}), and if it does, certifies the proposal for execution.
Once complete, the certification procedure typically moves the proposal to a \stress{Timelock} contract~\cite{Governance@Compound}, where it is actually executed.
We find that all DAOs with on-chain voting allow anybody to call the function to certify and execute the referendum.

Since on-chain mechanisms (i.e., voting, certification, and execution) take place without any intermediaries, DAOs using on-chain mechanisms typically add an \stress{\textbf{Execution Delay}}, a period between certification and execution.
The Execution Delay allows users who are unhappy with the proposal to divest themselves of their stake in the protocol, before the proposal (eventually) executes.
Note that proposals can be certified but not executed, as proposals can be \stress{cancelled} until they are executed (e.g. due to unexpected change in financial markets). 9 proposals in AAVE have been cancelled, for instance, \stress{after} certification \cite{aave-certified-canceled}.

\paraib{Centralized Certifier and Executor}\label{sec:centralized-cerifier-executor}
Some DAOs allow \stress{only} some ``trusted'' authority to certify and execute a referendum.
As we covered in \S\ref{s:off-chain-voting} all off-chain voting DAOs, by design, require a trusted entity to certify and execute a passed referendum on-chain.
This trusted entity often takes the form of a multi-signature (or multi-sig) wallet, with the signatories typically being the ``well-known'' stakeholders of the DAO.
Since \stress{only} the trusted entity can certify and execute proposals, it can essentially \stress{cancel} any proposal (even ones that passed) by simply not certifying and executing the proposal.
SuperRare holds elections to regularly elect signatories to their multi-signature wallet.
In all off-chain voting DAOs (except CowSwap and SuperRare), the certification and execution of a referendum are carried out in a \stress{single} transaction.
While DAOs using on-chain voting can assign this privilege to any wallet address, they do not do so in practice: Only one DAO with on-chain voting, Nexus Mutual, has a certification authority.

\paraib{Optimistic certification}
One approach to rid of the centralization concerns introduced by the need for a trusted entity in off-chain voting DAOs is \newterm{optimistic} certification.
In this mechanism, \stress{anybody} can \stress{assert} the correctness of off-chain voting results.
If this assertion is challenged, then a \stress{court} examines the voting and rules on the outcome.
Assertions and challenges are backed by collateral, and the losing part forfeits their collateral.
While the court could constitute any DAO user, they are composed of token holders of the protocol that developed the Optimistic Certification.
This approach obviates the need for a trusted entity for certification, and Kleros~\cite{kleros} and Optimistic Snap (oSnap)~\cite{osnap@medium} are two examples.
oSnap was proposed by Universal Market Access (UMA)~\cite{uma-osnap} and is used by Cow Protocol and Superrare~\cite{cow-osnap, superrare@osnap}.

\begin{mdframed}[style=Takeaways]
\takeaways{Trust and Transparency trade-offs}
Since DAOs with off-chain voting require a trusted party for executing proposals on-chain, we observe that almost all off-chain voting protocols use a multi-sig wallet for certification and execution.
Although optimistic certification counters centralization of decision-making power, its reliance on stakeholders of a different protocol to settle a dispute has crucial implications for their autonomy.
Rather than relying on a centralized multi-sig wallet belonging to the members of the DAO, the certification privilege is accorded to stakeholders of a different protocol;
the rationale behind the approach is that there is little overlap between the two protocols and the stakeholders of the other protocol will rationally determine the certification of the referendum.

In contrast, on-chain voting protocols allow anybody to certify and execute, since they rely on smart contracts for these functionalities; the smart contract checks if the proposal meets the requirements, and, if it does, it certifies and then executes the proposal. 
This provides a higher level of decentralization and autonomy in decision making, since the rules for governance are clearly written in code and will be executed based on the pre-defined rules.
However, such DAOs have a larger attack surface in regards to security \cite{feichtinger@sokattacksondao}, because even malicious proposals \stress{will} be executed if it adheres to the rules.
Execution Delays protect \stress{users} of a protocol from the DAO, which \stress{governs} the protocol. 
Since DAOs can make arbitrary changes to the existing protocol, the delay allows users, who disagree with the change, to remove their assets from the protocol before the proposal is executed.  

However, we find that 9 DAOs with on-chain voting have no execution delays, allowing proposals to be executed immediately after it has passed, making users of the protocol susceptible to an attack from the DAO.
The remaining on-chain DAOs have an execution delay of at least 1 day with a maximum delay of 3 days. 
Superrare also requires a KYC for candidates in order to stand up for election to become signatories of their multi-sig wallet \cite{ superrare@snapshotkyc}.
Given a multi-sig wallet's importance in the DAO, a KYC allows DAOs to prosecute signatories that do malicious actions.

\end{mdframed}

\subsection{Veto}\label{sec:canceler}
A \stress{vetoer} has the authority to veto or cancel referendums before they are executed.
Since it takes a long time for a proposal to be executed (from the time it was initiated), changes in market conditions during this time may necessitate canceling a proposal, 
since the stakeholders might vote differently now (on the same proposal) given the changes.
In our study, 38 DAOs allow for canceling a referendum after it has been proposed.

\paraib{Referendum Initiator}
\label{sec:proposer-of-proposal}
Some DAOs allow the referendum initiator to cancel the referendum; logically, it is an easily defensible design choice.
A referendum initiated by a stakeholder with sufficient voting power but little technical expertise may include bugs in the implementation.
If the initiator vetoes the proposal, but the DAO is still interested in the change brought about by the referendum, another stakeholder (perhaps with relatively more technical expertise) may fix the bugs and re-initiate the proposal.
Generally, anybody can cancel a proposal if the initiator's voting power falls below the proposal threshold~\S\ref{s:proposal-threshold}.
In our study, 9 DAOs with on-chain voting assign veto power to the referendum initiator.

\paraib{Veto Authority}
\label{sec:canceler-role}
Some protocols bestow the vetoing power to a specific wallet address, which typically is a multi-sig wallet. 
The rationale behind such a design is to assign explicitly a stakeholder or entity that is responsible for canceling potentially malicious or faulty proposals.
In our study, 11 on-chain protocols assign veto power to a veto authority.

\begin{mdframed}[style=Takeaways]
\takeaways{Trust and Transparency trade-offs}
Veto power is crucial for DAOs with on-chain voting, where referendums, unless vetoed, are automatically executed.
DAOs with off-chain voting  typically do \stress{not} need an explicit vetoer: Since the execution relies on a trusted entity, that entity can veto a proposal by simply not executing it.
On-chain voting DAOs in fact \stress{need} veto mechanisms, especially because they allow anybody to certify and execute a referendum (\S~\ref{sec:decentralized-cerifier-executor}).
Such mechanism prevents two important vulnerabilities. 
Firstly, there can be a faulty proposal, which \stress{could} be executed if the bug is discovered after the vote has passed.
Secondly, a malicious user can potentially buy a large amount of governance tokens, initiate malicious proposals, vote and then immediately sell their governance tokens, facing little to no economic downside. 
If the governance contract does not allow for any kind of veto, such referendums must be defeated every time at the cost of voters.

While a veto authority protects the DAO against malicious proposals, it also centralizes the DAO, since any arbitrary proposal can be cancelled.
Venus, for example, had a majority of registered votes in favor of the proposal to gain control of the DAO's treasury~\cite{venus-attack}.
However, this proposal was canceled by the DAO's vetoer. 
As a result, the DAO's treasury remained intact, although the choice of the majority (of votes) was rejected.
There are numerous examples of DAOs assigning vetoer roles in response to or anticipation of a potential attack~\cite{ensNovote, ens-veto-forum, comp304, indexedfinanceattack}.  
Giving the referendum initiator veto power may \stress{seem} harmless, but it allows them to cancel a proposal arbitrarily, forcing others to re-propose and re-vote, wasting time and money.

\end{mdframed}

%% file: 05-real-world-attacks.tex
\section{Case Studies: Governance Attacks}\label{s:real-world-attacks}

Thus far, we identified the key governance mechanisms and the trust assumptions inherent in every design choice.
Below, we review real-world attacks on governance and identify feasible strategies to prevent them.

\subsection{Identifying Governance Attacks}

Feichtinger \ea analyzes attacks on DAOs and identifies 28 incidents~\cite{feichtinger@sokattacksondao}.
Of these incidents, we label 16 as \stress{Governance Attacks}, based on the determination that a change in the \stress{governance mechanism} would have prevented the attack.
The rest resulted from of poor design of the underlying DeFi protocol (e.g., protocol vulnerability), lack of due diligence from the DAO (e.g., spam attack or proposal obfuscation), or could not be eliminated in any governance framework (e.g., bribing and vote buying).
6 out of the 16 governance attacks resulted from a bug in the governance contract and 10 were \stress{token acquisition} attacks.
Of the 10 token acquisition attacks, 5 attacks \stress{failed}.
In each case, a trusted entities could veto or refuse to certify proposals, which highlights the dire need for oversight in a DAO to protect it against malicious proposals.
One successful attack (Yuan Finance~\cite{yuan-finance-attack}) even had a vetoer, which did not cancel the malicious proposal.
Hence, vetoers must be aware of all ongoing referendums and identify malicious ones before they execute.

\subsection{Compound DAO attack}\label{s:compound-case-study}

\begin{figure}[t!]
    \centering
    \includegraphics[height=1.2in]{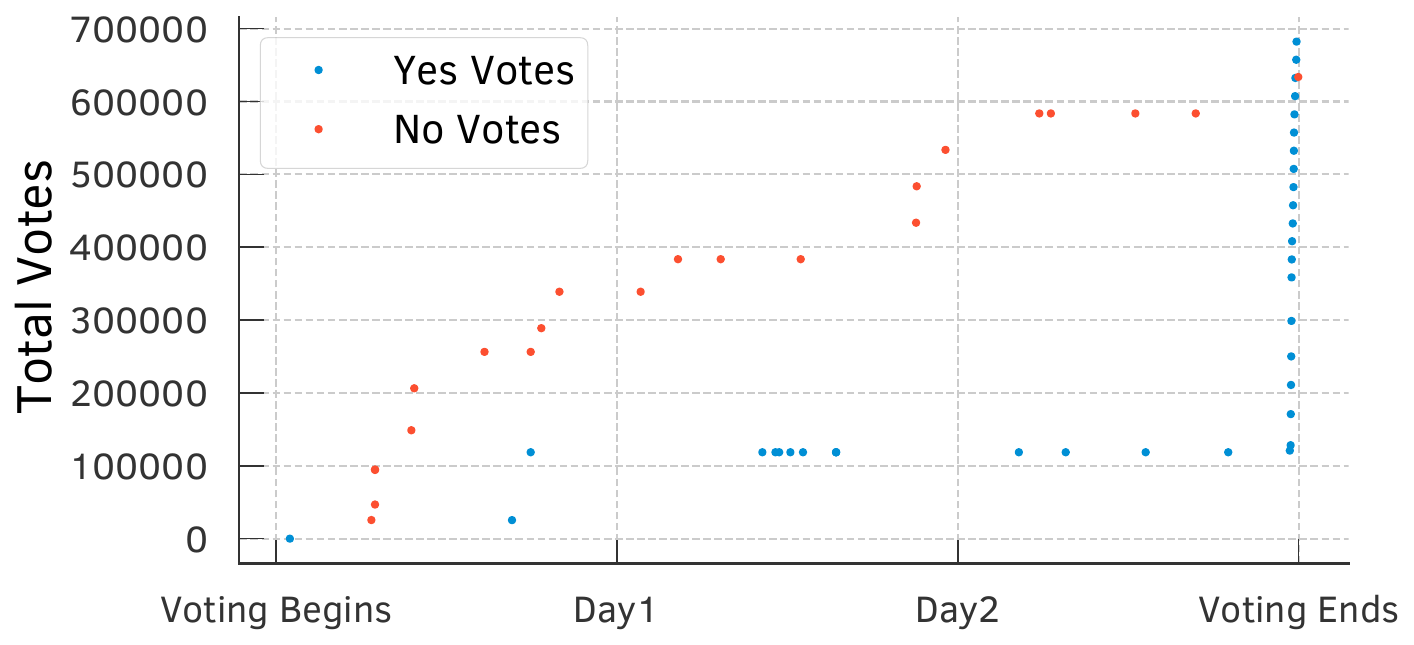}
    \figcap{Voting Tally of the malicious proposal in Compound. We see that the number of votes in favor sharply increases shortly before the end of the voting period.}
    \label{fig:compound-voting-anomalies}
\end{figure}

We now review the security vulnerabilities that were exploited in an attack on the Compound governance protocol in July 2024. 
Compound had three flaws that were exploited.

\paraib{Vote Buying}
In Compound, an entity can simply buy a large amount of tokens and immediately receive voting power. 
In total, there were in total 29 malicious wallet addresses that purchased a large amount of governance tokens in the previous 4 months~\cite{comp289}.
563,790 tokens were transferred to these addresses through four CEXes Centralized Exchanges (CEXes)---292,570 tokens through KuCoin~\cite{Kucoin}, 230,335 through ByBit~\cite{bybit}, 33,992 through HTX~\cite{htx} and 4,391 through OKX~\cite{okx}. A further 118,089 tokens were borrowed through the Compound Protocol.
Although the attackers held over 680,000 tokens, in the governance forums, the attacker's expected holding was suspected to be around 325,000~\cite{comp-forum-attack}.

\paraib{Voter Apathy and Vote Sniping} There were three malicious referendums in total with less votes cast against in each subsequent proposal~\cite{comp247,comp279,comp289}.
Since, the first two referendums failed, voters likely expected the third referendum to fail as well, and did not bother to vote against the proposal.
563,591 votes out of 682,191 (82\% of votes for) was cast within 170 blocks (or 34 minutes) before the end of the voting period, with the final vote being cast 40 blocks (or 8 minutes) before the end of the voting period~\cite{comp289} (as per \autoref{fig:compound-voting-anomalies}). This suddenly changed the fate of the referendum to pass. This is also known as \stress{vote sniping}.
The referendum narrowly passed by less than 50,000 votes. If measures mentioned in~\S\ref{s:voting-period} were taken, such as only allowing votes against a proposal at the end of the voting period, or extending the voting period if quorum was reached late, this attack would likely have failed, since it would have allowed other voters to react.

\paraib{Lack of oversight} There was no established oversight, such as a vetoer.
Due to the clear vote sniping that took place coupled with the strong dislike of the proposal within the Compound Governance Forums~\cite{compound-forum, comp-forum-attack}, a vetoer would have been justified in cancelling such a proposal.
Furthermore, having a vetoer would have disincentivized the attackers to even initiate such a referendum.
In fact, as a result of the attack, Compound governance passed a referendum to assign a vetoer to circumvent such future attacks~\cite{comp304}.

\subsection{On the Generality of Vulnerabilities} 
The vulnerabilities that were exploited in recent governance attacks \cite{comp-forum-attack,feichtinger@sokattacksondao} still exist across many of the DAOs in our study.
Since a large number of governance tokens can be easily bought in public markets, these design choices make DAOs susceptible to governance attacks via vote buying.
At the same time, we observe that excluding 4 DAOs---Lido, Frax, Origin, Angle---all DAOs use governance contracts that do \textit{not} prevent vote sniping. 
Besides, 15 DAOs in our study do \textit{not} have \stress{any} oversight (e.g. certifier, vetoer) to prevent malicious referendums.
Even worse, we find 7 DAOs---Uniswap, Radicle, Gitcoin, Silo, Ampleforth, Hop, and Cryptex--- vulnerable to a similar governance attack since their design choices have all the three inherent flaws that we discussed above.

%% file: 06-related-work.tex
\section{Related work}\label{s:related}

Regular updates are essential to meet the evolving needs of decentralized blockchain protocols. 
Consensus among stakeholders is crucial for these changes, and is often achieved through social norms rather than formal voting.
For example, Bitcoin and Ethereum use Bitcoin Improvement Proposals (BIPs) and Ethereum Improvement Proposals (EIPs), respectively, where users, developers and researchers engage in community discussions to determine if a proposal gains enough support for the core developers to implement~\cite{bips,eips}.
However, deep divisions can cause a blockchain to split, reducing its value and security, such as Bitcoin being split into Bitcoin and Bitcoin Cash~\cite{verge-bitcoin, coindesk-bitcoin-cash}; and Ethereum into Ethereum and Ethereum Classic~\cite{kiffer2017hotnets, coindesk-ethereum-classic}.
Fracassi \ea showed that EIP proposals and implementations of these proposals are highly centralized~\cite{ethereum-centralization}.

In a separate paradigm, the governance of blockchain systems has been described to be analogous to social contract theory~\cite{reijers2016governance} or even corporate governance~\cite{allen@blockchaingov}, where consensus is achieved through formal voting. 
Filippi and Mcmullen~\cite{defilippi@hal} highlight two types of governance in blockchain systems based on where they are operationalized: on-chain and off-chain governance. On-chain governance has fixed, enforceable rules but struggles with unexpected situations, while off-chain governance is flexible and adaptable but harder to enforce.
This study focuses on formal voting procedures in DAOs, examining design choices of governance contracts and their implications for decentralization, autonomy, and security.

Although DAOs are intended to be decentralized and autonomous, prior work shows significant deviations from these ideals.
Recent studies showed that voting power in DeFi projects is very centralized, questioning whether the protocols are truly decentralized~\cite{fritsch@2022votingpower, messias2023gov, feichtinger2023hidden, kitzler2023governance}. 
Sharma \ea studied how 10 DAOs work in practice, and looked at their degrees of decentralization and autonomy~\cite{sharma2023unpacking}.
Kiayias and Lazos derived seven fundamental properties of blockchain governance from 10 widely used blockchain platforms, finding that all of them have some deficiency in their governance~\cite{kiayias@2022governance}. 
Feichtilinger \ea also did an extensive survey on governance attacks on DAOs, and classified each attack along different dimensions~\cite{feichtinger@sokattacksondao}.
The work highlights the different ways that DAOs can potentially fail, ranging from buggy smart contracts to social engineering.
Despite focusing on smart contract bugs and DAO attacks, these works largely overlook how governance contract design choices contribute to various risks. We address this gap by showing how fundamental design decisions lead to centralization and security vulnerabilities across multiple dimensions.

Most relevant to this work, Tan \ea posed several open problems relevant to DAOs from various perspectives (e.g. computer science, economics, law)~\cite{tan2023open}.
While the work identifies several unsolved problems, they overlook DAO governance in practice, the various design decisions and their trade-offs.
Introducing a new measure for centralization of DAOs, Austgen \ea highlights the impact of voter apathy, delegation, and other factors~\cite{austgen2023dao}.
While they focus mainly on distribution of voting power, our work takes a holistic view of DAO governance, examining how a DAO's governance contract has implications on how governance is exercised. 
In our extensive study of \surveysz{} DAOs, by breaking DAO governance into granular components, we uncover design trade-offs, and offer actionable insights for researchers and practitioners for better realization of DAOs.

%% file: 07-conclusion.tex
\section{Concluding Remarks}\label{s:conclusion}

In realizing a \ac{dao}, numerous design choices must be made, each of which must balance complex trade-offs between efficiency, security, and fairness (or the decentralized nature) of the \ac{dao}.
The introduction of a vetoer, for instance, improves security but at the cost of eroding decentralization; off-chain voting encourages participation, but requires a centralized party for executing the vote.
In this work, we investigated how these fundamental design ``knobs'' affect various security, fairness, and performance considerations, and we presented an examination of the \acp{dao} ``in theory.''
Said differently, while prior work~\cite{feichtinger@sokattacksondao} looked at vulnerabilities that arise in the implementation of these knobs, our work presents the fundamental trade-offs \stress{inherent} in the design.
We also discover governance attacks, a distinct class of vulnerabilities, that exploit the governance mechanisms of \acp{dao}, regardless of their implementations.

%% file: appendix.tex

\input{98b-compound-analysis}
\input{98a-chain-address}

%% file: 98b-compound-analysis.tex
\subsection{Compound Case study}

\paraib{Attack Timeline}\label{s:attack-timeline} \label{sec:compound-attack-timeline}
On May 6th, 2024, a proposal was put forth by Humpy~\cite{humpy-address}, a known DeFi whale, in the Compound DAO, Prop. \#247~\cite{comp247}, which transferred 92,000 COMP tokens to the a Multisig wallet (called GOLD Multisig) ~\cite{goldmultisig-contract}. 
This proposal was rejected with 710,978 voting against and only 95,865 votes for (Humpy was the only large voter to vote for the proposal).
On July 15th, a proposal, Prop. \#279 ~\cite{comp279}, was put forth by Humpy again, where the 92,000 COMP tokens are transferred to a goldCOMP smart contract.
The proposal was again flagged on the Governance Forum ~\cite{comp-forum-attack}. 
The goldCOMP smart contract is controlled by the same Multisig, and this was seen by the Compound community as an attack on Governance.
This proposal was rejected by the users again with 578,664 voting against and 118,530 voting for.
On July 24, 2024, a proposal was put forth in the Compound DAO transferring 499,000 COMP tokens to goldCOMP smart contract (Prop. \#289) ~\cite{comp289}.
This was a significantly large amount of tokens than the previous 2 proposals.
The proposal was again flagged on the Governance Forum. 
However, this proposal passed narrowly with 682,191 voting for and 633,636 voting against.

\begin{figure}[th]
    \centering
    \includegraphics[height=1.2in]{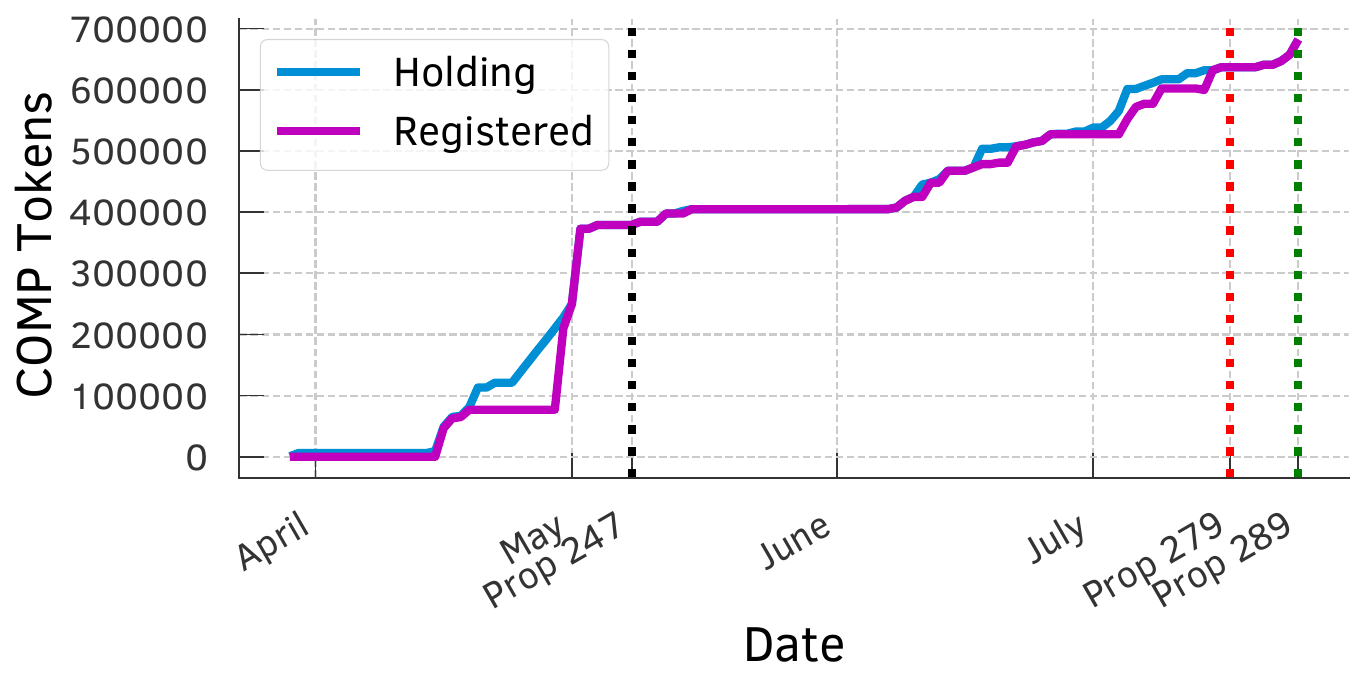}
    \figcap{Voting Tally of the malicious proposal in Compound. We see that the number of votes in favor sharply increases shortly before the end of the voting period.}
    \label{fig:compound-voting-anomalies-2}
\end{figure}

\paraib{Voting Anomalies.}\label{s:voting-anomalies}
There are anomalies in the voting pattern and voters of Prop. \#289.
Firstly, 563,591 votes out of 682,191 (82\% of votes for) was cast within 170 blocks (or 34 minutes) before the end of the voting period. The final vote was cast 40 blocks (or 8 minutes) before the end of the voting period.
This clearly constitutes \stress{vote sniping}
and, does not allow other voters to respond to the change in the outcome of the vote.
Furthermore, some voters might not have voted because of the cost to voting and the proposal was already headed towards their wanted outcome.

Since, the previous proposal (Prop. \#279) ended with only 118,530 votes for, users might have not have voted on this proposal because they incorrectly assumed that more votes would not be cast.
Analyzing the voters that voted for Prop. \#289, we identified 29 addresses in total that either voted or were delegators.
On March 29, 2024, we see the first significant transfer of COMP tokens to the Humpy address. Before this period, the total amount of COMP tokens held in these addresses was 853 COMP tokens. In the next 4 month, i.e. until late July, they accumulated a total of 682,181 COMP tokens. 
Before this period, none of these addresses had ever participated in Compound governance, and had no delegated votes.
There were two sources through which the addresses accumulated COMP tokens.
The first is through the Compound Protocol, where Humpy borrowed 118,089 COMP tokens.
The second is through Centralized Exchanges. There were a total of 292,570 COMP tokens transferred through KuCoin ~\cite{Kucoin}, 230,335 COMP tokens transferred through ByBit ~\cite{bybit}, and 33,992 COMP tokens through HTX ~\cite{htx} and 4,391 COMP tokens transferred through OKX ~\cite{okx}. This accounts for 563,790 COMP tokens in total transferred to these addresses through four centralized exchanges.  
Furthermore, all addresses (except for Humpy who voted on Prop. \#277, \#287, \# 288), voted only on Prop. \#289.

\paraib{Aftermath and Solution}\label{s:attack-aftermath-solution}
In response to the attack, a proposal was put forth to transfer the admin of the Timelock to a Multi-Signature Wallet, thereby giving the wallet the right to propose and pass any referendum~\cite{comp290}.
However, the malicious proposal itself could not be stopped from being certified or executed, since the Compound did not have any certifiers or vetoers.
In light of this, a compromise was reached with the Compound DAO, where both proposals would be canceled.
In return, Compound would roll out a staking program, where Compound token holder can stake their tokens in order to receive a portion of the profits made by the protocol ~\cite{comp-forum-staking}.
Furthermore, on August 10th (Prop. \#304) ~\cite{comp304}, a proposal was put forward to add a vetoer, which is able to cancel malicious proposals that have passed and are awaiting execution. 
The proposal passed with 1,398,783 votes for and no votes against.\footnote{None of the addresses associated with the attack voted on this proposal.}
This again highlights the fundamental problem of balancing decentralization with checks and balances within a DAO. 
The fundamental problem with the Compound DAO was that there were no gatekeepers to oversee and interject when there are potentially malicious activities.
Hence, this is remedied this by adding a vetoer.
Even though gatekeepers centralize the decision making process in a DAO, they provide an important check against malicious proposals.

7 DAOs are susceptible to a similar attack as was carried out on Compound : Uniswap, Radicle, Gitcoin, Silo, Ampleforth, Hop, Cryptex. 
Each DAO uses Token Holding, and has a large amount of governance tokens available to be bought in public markets (e.g. CEX, DEX).
They also lack any mechanism to prevent vote sniping, and lacks any oversight such as a vetoer.

%% file: 98a-chain-address.tex
\subsection{On-chain Contract Addresses}\label{sec:on-chain-contract-addresses}

In this section, we present the list of all relevant contract addresses that were analyzed in our study, as presented in~\autoref{tab:on-chain-adds}. This list includes contract addresses obtained from both the Ethereum blockchain, which contains contracts deployed on-chain, and the Snapshot platform, which is used for governance contracts that employ off-chain voting mechanisms.

\begin{table*}[btp]
\small
\tabcap{%
On-chain and snapshot addresses
}\label{tab:on-chain-adds}%
\centering
\resizebox{0.92\textwidth}{!}{%
\begin{tabular}{@{}r|c|c@{}}
\toprule
\thead{Protocol} & \thead{On-chain Address}  &  \thead{Off-chain space}  \\
\midrule

Optimism & 	
0xcDF27F107725988f2261Ce2256bDfCdE8B382B10 (on Optimisim) & x \\

Arbitrum & 0xf07DeD9dC292157749B6Fd268E37DF6EA38395B9 (Core) & arbitrumfoundation.eth \\
& 0x789fC99093B09aD01C34DC7251D0C89ce743e5a4 (Treasury) (On Arbitrum) & \\

Uniswap & 0x408ED6354d4973f66138C91495F2f2FCbd8724C3 & uniswapgovernance.eth  \\

ENS & 0x323A76393544d5ecca80cd6ef2A560C6a395b7E3 & ens.eth \\
 
Maker & 0x0a3f6849f78076aefaDf113F5BED87720274dDC0 & x \\

Lido & 0xF0211b7660680B49De1A7E9f25C65660F0a13Fea (Easy Track) & lido-snapshot.eth \\

 & 0x2e59A20f205bB85a89C53f1936454680651E618e (Governance) & \\

Frax Finance & 0x953791D7C5ac8Ce5fb23BBBF88963DA37a95FE7a (Omega) & frax.eth, fpis.eth \\

& 0xd74034C6109A23B6c7657144cAcBbBB82BDCB00E (Alpha) &  \\

AAVE & 0xEC568fffba86c094cf06b22134B23074DFE2252c & aave.eth  \\

Compound & 0xc0Da02939E1441F497fd74F78cE7Decb17B66529 & x \\

Radicle & 0x690e775361AD66D1c4A25d89da9fCd639F5198eD & gov.radworks.eth \\

0x Protocol & 0x0bB1810061C2f5b2088054eE184E6C79e1591101 & 0xgov.eth \\

Gitcoin & 0x9D4C63565D5618310271bF3F3c01b2954C1D1639 & gitcoindao.eth \\

Silo Finance & 0xA89163F7B2D68A8fbA6Ca36BEEd32Bd4f3EeAf61 & silofinance.eth \\

Lyra & 0xe8642cc1249F08756e70Bb8eb4BE0e6c09254fed  & lyra.eth  \\

API3 & 0xDb6C812E439ce5c740570578681ea7aaDba5170b (Primary) &  \\
  & 0x1c8058E72E4902B3431Ef057E8d9a58A73F26372 (Secondary) & \\
  & 0x05811Ad31cbD5905e4e1427482713E3fb04A4c05 (Old Voting) & \\

Ampleforth & 0x8a994C6F55Be1fD2B4d0dc3B8f8F7D4E3a2dA8F1 & 	
ampleforthorg.eth \\

Instadapp & 0x0204Cd037B2ec03605CFdFe482D8e257C765fA1B & instadapp-gov.eth \\

Rari & 0x6552C8fb228f7776Fc0e4056AA217c139D4baDa1 & x  \\

NounsDAO & 0x6f3E6272A167e8AcCb32072d08E0957F9c79223d & x \\

Curve  & 0xE478de485ad2fe566d49342Cbd03E49ed7DB3356 (Ownership) & x \\
& 0xBCfF8B0b9419b9A88c44546519b1e909cF330399 (Parameter) & \\

Origin & 0x1D3Fbd4d129Ddd2372EA85c5Fa00b2682081c9EC (Current) & x \\
& 0x3cdD07c16614059e66344a7b579DAB4f9516C0b6 (Third version) & \\
& 0x72426BA137DEC62657306b12B1E869d43FeC6eC7 (Second Version) & \\
& 0x8a5fF78BFe0de04F5dc1B57d2e1095bE697Be76E (First Version) & \\

Hop DAO & 0xed8bdb5895b8b7f9fdb3c087628fd8410e853d48 & hop.eth \\

Cryptex & 0x874C5D592AfC6803c3DD60d6442357879F196d5b & cryptexdao.eth \\

Nexus Mutual & 0x4A5C681dDC32acC6ccA51ac17e9d461e6be87900 & community.nexusmutual.eth \\

Mantle & 0x78605Df79524164911C144801f41e9811B7DB73D & bitdao.eth \\

Gnosis  & 0x849D52316331967b6fF1198e5E32A0eB168D039d & gnosis.eth \\

SuperRare & 0x860a80d33E85e97888F1f0C75c6e5BBD60b48DA9 & 	
superraredao.eth \\


Aevo & 0xDAEada3d210D2f45874724BeEa03C7d4BBD41674 & rbn.eth \\

Research Hub Foundation & 0x5222FF25F4DFC02d173C2cbd2055EE1D35f291F1 (Treasury 1) & researchhub.eth \\

& 0xE3648e99B6E68A09e28428790D12B357f081dBe0 (Treasury 2) & \\

& 0xC4cfa2BdAE08416312fAa0B72758E1F3750f81e3 (Treasury 3) & \\

Stargate Finance & 0x65bb797c2B9830d891D87288F029ed8dACc19705 & stgdao.eth \\

Uma & 0x7b292034084A41B9D441B71b6E3557Edd0463fa8 & uma.eth \\

Illuvium & 0x58C37A622cdf8aCe54d8b25c58223f61d0d738aA  & ilvgov.eth, ilv.eth \\

Cowswap & 0xcA771eda0c70aA7d053aB1B25004559B918FE662 & cow.eth  \\ 

Goldfinch & 0xBEb28978B2c755155f20fd3d09Cb37e300A6981f & goldfinch.eth \\

Sturdy Finance & 0xfE6DE700427cc0f964aa6cE15dF2bB56C7eFDD60 & sturdyfi.eth  \\

Euler & 0xcAD001c30E96765aC90307669d578219D4fb1DCe  & eulerdao.eth \\

SAFE & 0x1d4F25bC16b68C50B78E1040BC430a8097Fd6f45 & safe.eth \\

JPEG'd & 0x51C2cEF9efa48e08557A361B52DB34061c025a1B  & jpeg’d.eth \\

Tokenlon & 0x3557BD3d422300198719710Cc3f00194E1c20A46 & tokenlon.eth \\

Botto & 0x35bb964878d7B6ddFA69cF0b97EE63fa3C9d9b49 & botto.eth \\

Balancer & 0x10A19e7eE7d7F8a52822f6817de8ea18204F2e4f & balancer.eth \\

Galxe & 0x03a42D37066726FfDc8F2BCd1C422f72A1717882 & gal.eth \\

Synthetix & 0xE832C302D1160EAe57045eb9d9Ea14daBd2E229c (Optimism Election) & https://gov.synthetix.io/\#/ \\

& 0xEb3107117FEAd7de89Cd14D463D340A2E6917769 (Protocol Dao) &  \\

Sushiswap & 0xe94B5EEC1fA96CEecbD33EF5Baa8d00E4493F4f3 & sushigov.eth \\

Gearbox & 0x7b065Fcb0760dF0CEA8CFd144e08554F3CeA73D1 & gearbox.eth \\

Paraswap & 0x5A61D9214adEFD7669428a03A4e8734A00E9F464 & paraswap-dao.eth \\

ParagonsDAO & 0xc977CBadD359aE06b236D9581e37fd5A03E54b84 & paragonscouncil.eth, paragonsdao.eth \\

Alchemix & 0x8392F6669292fA56123F71949B52d883aE57e225  & alchemixstakers.eth  \\

1Inch & 0x7951c7ef839e26F63DA87a42C9a87986507f1c07 & 1inch.eth \\

Angle Protocol & 0xdC4e6DFe07EFCa50a197DF15D9200883eF4Eb1c8 & anglegovernance.eth \\

Shutter DAO 0x36 & 0x36bD3044ab68f600f6d3e081056F34f2a58432c4 & shutterdao0x36.eth \\

Yearn Finance & 0xFEB4acf3df3cDEA7399794D0869ef76A6EfAff52 & veyfi.eth  \\

Shapeshift & 0x90A48D5CF7343B08dA12E067680B4C6dbfE551Be & shapeshiftdao.eth \\

Decentraland & 0x41E83d829459F99Bf4Ee2E26D0D79748Fb16b94F & 	
snapshot.dcl.eth \\

\bottomrule
\end{tabular}}


\label{tab:heuristic1}\vspace{-6pt}

\end{table*}